\documentclass[reprint,amsmath,amssymb,aps]{revtex4-1}

\usepackage{multirow}
\usepackage{braket}
\usepackage{graphicx}
\usepackage{dcolumn}
\usepackage{bm}
\usepackage{array}
\usepackage{color}
\usepackage{lineno}

\def\bfk{{\bf k}}
\def\bfq{{\bf q}}
\def\bfR{{\bf R}}
\def\bfr{{\bf r}}
\def\bfT{{\bf T}}
\def\chipm{\chi^{+-}(\bfq,\omega)}
\def\chipm{R^{+-}(\bfq,\omega)}
\def\ef{E_{\rm Fermi}}
\def\mb{\mu_{\rm B}}
\def\dex{{\Delta}E_{\rm x}}

\def\ham{\smH^{\rm LDA}_0}
\def\hamz{\smH_0}
\def\imko{-{\rm Im}[K^{+-}(\bfq=0,\omega)]}
\def\jij{{J_{ij}}}
\def\jk{{J({\bf q})}}
\def\lsmo{\rm {{La_{0.7}Sr_{0.3}MnO_3}}}

\def\ud{{\uparrow\downarrow}}
\def\SWE{{\omega({\bf q})}}

\def\ImRpm{{\rm Im}[R^{+-}(\bfq,\omega)]}

\newcommand{\req}[1]{\mbox{Eq. (\ref{#1})}}
\newcommand{\retable}[1]{\mbox {Table \ref{#1}}}
\newcommand{\refig}[1]{\mbox{Fig. \ref{#1}}}
\newcommand{\refigs}[1]{\mbox{Figs. \ref{#1}}}

\def\nbvk{N}

\def\dxy{d_{xy}}

\def\Psiknr{\Psi_{\bfk n}(\bfr)}

\def\ekn{{\varepsilon_{\bfk n}}}

\def\wirRt{w_{\bfR i}(\bfr-\bfR-\bfT)}
\def\wjrRtp{w_{\bfR' j}(\bfr-\bfR'-\bfT')}

\def\wkir{w_{\bfR{i}}^{\bfk}(\bfr)}
\def\akni{a_{\bfR i}^{\bfk n}}
\def\aknj{a_{\bfR' j}^{\bfk n}}

\def\akbj{a^{\bfk n*}_{\bfR{j}\beta}}
\def\akbl{a^{\bfk n}_{\bfR'{l}\beta}}
\def\akqai{a^{\bfk+\bfq n'}_{\bfR{i}\alpha}}
\def\akqak{a^{\bfk+\bfq n'*}_{\bfR'{k}\alpha}}
\def\aqijkl{\akbj\,\akbl\,\akqai\,\akqak}
\def\smH{\mathcal H}

\begin{document}

\title{Spin wave dispersion of 3$d$ ferromagnets based on QSGW calculations}

\author{H. Okumura}
\email{okumura.haruki@mat.eng.osaka-u.ac.jp}
\affiliation{
  Division of Materials and Manufacturing Science, Graduate School of Engineering, Osaka University, Osaka, Japan.
}

\author{K. Sato}
\affiliation{
  Division of Materials and Manufacturing Science,
  Graduate School of Engineering, Osaka University,
  Osaka, Japan.
}
\affiliation{
   Center for spintronics research network (CSRN), Osaka University,
   Osaka, Japan.
 }

\author{T. Kotani}
\affiliation{
  Department of Applied Mathematics and Physics, Tottori University,
  Tottori, Japan.
}
\begin{abstract}
We calculate transverse spin susceptibility in the linear response method based on
the ground states determined in the quasi-particle self-consistent $GW$ (QSGW) method.
Then we extract spin wave (SW) dispersions from the susceptibility.
We treat bcc Fe, hcp Co, fcc Ni, and B2-type FeCo.
Because of the better description of the independent-particle picture in QSGW, calculated spin stiffness constants for Fe, Co, and Ni give much better
agreement with experiments in QSGW than that in the local density approximation (LDA), where the stiffness for Ni in LDA is two times bigger than the experiment.
For Co, both acoustic and optical branches of SWs agree with the experiment.
As for FeCo, we have some discrrepancy between the spin stiffness in QSGW and that in the experiment.
We may need further theoretical and experimental investigations on the discrepancy.
\end{abstract}
\maketitle

\section{Introduction}
Spin wave (SW) is one of the important factors to control magnetic properties of material.
SW is excited at considerably low temperature compared to room temperature (RT), and its energy range typically lies in a few hundred meV. 
When one magnetic moment tilted from the parallel spin configuration, 
the exchange interaction triggers the SW propagation throughout the
material as collective excitation. 
We can observe SWs in bulk materials by inelastic neutron scattering experiment,
{\it e.g.}, in bcc Fe \cite{Fe_expt1}, fcc Ni \cite{Ni_expt2}, and even half-metals 
like perovskite $\lsmo$ \cite{lsmo_expt}.
In addition to collective excitation,
another magnetic excitation like spin-flip excitation is called Stoner excitation,
whose excitation energy is related to the exchange splitting $\dex$.
We can experimentally observe Stoner excitation by the high energy experiment such as
spin-polarized electron energy loss spectroscopy (SPEELS) \cite{Stoner_Vollmer}.
High energy SWs are strongly damped because of the hybridization with the Stoner excitation.

Let us explain how we determine the spin stiffness $D$ experimentally.
From the macroscopic point of view,
the Bloch's {\it $T^{\frac{3}{2}}$} rule \cite{Bloch_rule}
in the temperature dependence of magnetization {\it M(T)} is derived from the SW theory.
For the wave vector $\bfq \sim0$, the SW dispersion $\SWE$ behaves as
$\SWE=D\bfq^2$.
Since this behavior of $\SWE$
results in the {\it $T^{\frac{3}{2}}$} rule in low temperature,
we can determine $D$ by analyzing the temperature dependence of magnetization \cite{Fe_expt2}.

We mainly have three methods to calculate $\SWE$ in the first-principles methods.
The first one is the Lichtenstein formula (LF) \cite{Lichtenstein}.
Assuming the Heisenberg model, we calculate exchange
interaction $\jij$ or its Fourier transform $\jk$ based on the magnetic force theorem \cite{MFT}. Here $i,j$ are for site 
indices. Then $\SWE$ is calculated from $\jk$.
In Ref. \onlinecite{Lichtenstein}, they calculated $\jij$ up to the second nearest neighbors, resulting in $D$, which are in good agreement with experiments for Fe and Ni. Later, Pajda {\it et al.} investigated the convergence of $D$ for a range of neighbors and found that converged $D$ are in good agreement with experiments 
for Fe but overestimated for Ni \cite{Pajda}.

The second one is the frozen magnon method (FMM) \cite{FMM_Halilov},
which assumes the Heisenberg model as in LF.
In FMM, we employ adiabatic approximation; namely, we neglect motions of the magnetic moment compared to electron motions.
Then we calculate $\jk$ from the constraint spin-spiral 
configurations with the fixed magnitude of the magnetic moment.
Once we get $\jk$, we solve the eigenvalue problem for deriving $\SWE$.
This method works well for bcc Fe \cite{FMM_Halilov, Grotheer}.
Note that we can not describe the decay of collective SWs (Stoner damping) in both of these two methods.

The third one is the linear response (LR) method for transverse spin susceptibility
$\chipm$ \cite{LR_Gross}. The LR method directly gives $\SWE$ in the reciprocal space.
Cooke {\it et al.} first introduced the LR method for calculating $\chipm$, and they discussed Stoner damping in SWs in bcc Fe and fcc Ni \cite{LR_Cooke}.
Savrasov treated spin fluctuations based on the many-body perturbation theory and reproduced the experimental $\SWE$ \cite{LR_Savrasov}.
Karlsson and Aryasetiawan also calculated $\chipm$
based on the Green function method \cite{LR_Karlsson}.
From a view of computational efficiency,
\c{S}a\c{s}{\i}o\v{g}lu {\it et al}. proposed a
LR method with maximally-localized Wannier function (MLWF) \cite{LR_Sasioglu}. 
In the method, we decrease to the second power of the number of a Wannier basis set and
we can decrease the calculation cost.
With this efficient method, they can use fine $\bfq$ mesh for calculating $\chipm$.

These three methods mainly have been applied to the ground states given in the
local density approximation (LDA). 
However, the ground state given in LDA is not necessarily good enough.
For example, Sponza {\it et al.} shows that $3d$-bandwidth and $\dex$ in LDA are not good enough to 
calculate $\SWE$ \cite{Sponza}.
In antiferromagnetic transition metal oxides such as NiO and MnO, 
the calculated $\SWE$ does not agree with the experiment
due to too small $\dex$ and too small bandgap \cite{TMO_QSGW}.
Serious disagreement is also found in the $\SWE$ in $\lsmo$, 
for which LDA fails to reproduce the half-metallic electronic structure of that compound \cite{LSMO_QSGW}.
It is possible to start from the ground states of LDA+$U$; however,
we sometimes have difficulty in determination of $U$.
It may suggest a limitation of LDA+$U$ itself.

To overcome such limitations in LDA, Kotani {\it et al.} calculated
$\SWE$ for strongly-correlated materials in an LR method
for the ground states determined in the quasi-particle self-consistent $GW$ (QSGW) method \cite{TMO_QSGW,LSMO_QSGW}.
Then we see reasonable agreement with experiments for NiO and MnO because QSGW gives good descriptions of the band quantities such as $\dex$ and bandgaps \cite{Deguchi_2016}.
We expect such good agreement for wide-range of materials.
However, Kotani's LR method used in Refs. \cite{TMO_QSGW,LSMO_QSGW}
is too simple to apply a wide range of materials.
Thus we implemented the efficient LR method to calculate 
$\chipm$ based on the MLWF given by \c{S}a\c{s}{\i}o\v{g}lu {\it et al.} \cite{LR_Sasioglu} 
in QSGW calculation package $ecalj$ compiled by Kotani {\it et al.} \cite{ecalj}.
We demonstrate how the method works for typical ferromagnets such as 
bcc Fe, fcc Ni, hcp Co, and B2 FeCo (CsCl structure) and we discuss the difference between LDA and QSGW.
Except for FeCo, the SWs in QSGW agree with experiments.
We find some discrepancies for FeCo.

\section{Computational methods}\label{sect_2}
\subsection{quasiparticle self-consistent {\it GW} (QSGW)}
Until now, varieties of $GW$ calculations 
based on the Hedin's GW approximation \cite{hedin_gw1,hedin_gw2} 
have been performed since it is introduced to the first-principles 
calculations by Hyberstein and Louie \cite{HybersteinLouie1986}.
Most of the $GW$ calculations are so-called one-shot $GW$.
Starting from $G^0$ for the one-body Hamiltonian in LDA $\ham$, 
we calculate corrections to the eigenvalues of $\ham$ to reproduce quasiparticle energies.
In the one-shot $GW$, the self-energy for the corrections is given as
$\Sigma(1,2)=iG^0(1,2)W(1^+,2)$, where we use notation $1\equiv(\bfr_1,t_1)$.
The screened Coulomb interaction $W(1^+,2)$ is calculated
as $W=(1-vP)^{-1}v$ from the bare Coulomb interaction $v$ and
the polarization function $P=-iG^0\times G^0$.
The one-shot $GW$ has a shortcoming
since the one-shot $GW$ is just a perturbation on top of $\ham$.

To overcome the shortcoming of the one-shot $GW$,
we utilize QSGW method \cite{qsgw1,qsgw2,qsgw3} implemented in $ecalj$ package \cite{ecalj}. 
Let us summarize QSGW method. 
At first, recall the above $GW$ procedure which can be applicable to any
static one-body Hamiltonian $\smH_0(\bfr,\bfr')$ as
\begin{eqnarray}
  \smH_0(\bfr,\bfr')=
  -\frac{\nabla^2}{2}+V_{\rm ext}+V_{\rm H}+V_{\rm xc}(\bfr,\bfr'),
  \label{eq:ham0}
\end{eqnarray}
where we have the external potential $V_{\rm ext}$, the
Hartree potential $V_{\rm H}$, and the non-local exchange-correlation potential $V_{\rm xc}(\bfr,\bfr')$. 
With $\Sigma(1,2)=iG^0(1,2)W(1^+,2)$ where $G^0=1/(\omega-\smH_0)$, 
we have the energy-dependent one-body Hamiltonian $\smH(\bfr,\bfr';\omega)$ as
\begin{eqnarray}
  \smH(\bfr,\bfr';\omega)=-\frac{\nabla^2}{2}+V_{\rm ext}+V_{\rm H}+\Sigma(\bfr,\bfr';\omega).
  \label{eq:ham}
\end{eqnarray}
That is, $GW$ approximation gives a procedure $\smH_0 \rightarrow \smH$.
QSGW requires ``quasiparticle self-consistency'', that is,
minimization of the difference between $\smH_0$ and $\smH$.
The minimization gives the procedure $\smH\rightarrow \smH_0$,
replacing the $\omega$-dependent $\Sigma$ in \req{eq:ham} with 
the static non-local exchange-correlation potential $V^{\rm xc}$ as
\begin{eqnarray}
  V^{\rm xc} =\frac{1}{2} \sum_{ij} \ket{\psi_i} \Bigl\{
    {\rm Re}\bigl[\Sigma(\varepsilon_i)\bigr]_{ij}+
    {\rm Re}[\Sigma(\varepsilon_j)]_{ij}
    \Bigr\} \bra{\psi_j},
\end{eqnarray}
where eigenvalues $\varepsilon_i$ and eigenfunctions $\psi_i$ are those of $\smH_0$.
This defines a procedure to give a new $\smH_0$, $\smH \rightarrow \smH_0$.
Thus we finally have a 'quasiparticle self-consistency' cycle 
$\smH_0\rightarrow \smH\rightarrow \smH_0\rightarrow \smH\rightarrow \cdots$ 
(or $G^0\rightarrow G\rightarrow G^0\rightarrow \cdots$) until converged.

\subsection{Dynamical magnetic susceptibility}\label{s_sect2}
In LR, we follow the procedure given in Ref. \cite{LR_Friedrich,LR_Sasioglu}.
Here we treat the transverse spin susceptibility $R^{+-}(1,2)$,
which describes the response of the expectation value of a spin density operator
$\hat{\sigma}^+(1)$ to the the external magnetic field $B^-(2)$ as,
\begin{eqnarray}
R^{+-}(1,2)
=\frac{\delta  \braket{\hat{\sigma}^+ (1)} }{\delta B^+ (2)}  \hspace{0.3cm},
\end{eqnarray}
where $1=(\bfr_1,t_1)$. See Eq. (20) in Ref. \onlinecite{LR_Friedrich}.
Here the expectation value of $\hat{\sigma}^+(1)$ is given as
\begin{eqnarray} 
\braket{\hat{\sigma}^+(1)}=-i\sum_{\alpha,\beta}\sigma^{+}_{\beta\alpha}G_{\alpha\beta}(1,1^{+})
\hspace{0.3cm} (\alpha,\beta \in \{\uparrow,\downarrow\}),
\end{eqnarray} 
where $G(1,1^+)$ is the single-particle Green function from 1 to $1^+$.
For our calculation below, it is convenient to consider
four-points representation $R^{(4)}_{\ud}(12,34)$.
The trace of matrix $R^{(4)}_{\ud}(11,33)$ leads to two-point representation $R^{+-}(1,2)$.

In order to obtain $R^{(4)}_{\ud}(12,34)$,
we solve the Bethe-Salpeter equation where we use the static screened
Coulomb interaction $W(1^+,2)$ which is $\propto \delta(t_1-t_2)$.
It is
\begin{eqnarray} 
&R^{(4)}_{\ud}(12,34)=K_{\ud}(12,34)\nonumber\\
&+\int\hspace{-0.2cm}\int K_{\ud}(12,56) \, W(5^+, 6) \, R_{\ud}(56,34) \, d5 d6,
\label{eq:R}
\end{eqnarray} 
where $K_{\ud}(12,34)$ is
the non-interacting two-particle (particle-hole with opposite spin) propagator given as
\begin{eqnarray}
-K_{\ud}(12,34)=-iG_{\uparrow}^0(1,3)G_{\downarrow}^0(4,2^+),
\label{eq:K}
\end{eqnarray}
where we consider $t_1=t_2$ and $t_3=t_4$, $i.e.$,
$K_{\ud}(\bfr_1,\bfr_2;\bfr_3,\bfr_4;t_1-t_3)$.
The Fourier transform is 
from $t_1-t_3$ to $\omega$.
We symbolically solve \req{eq:R} to be
$R=K+KWK+KWKWK+\cdots = K(1-WK)^{-1}$,
where the numerator $K$ describes the Stoner excitations,
whereas zeros of the denominator $(1-WK)$ gives the collective excitation.

This $K_{\ud}$ is given as
\begin{eqnarray} 
&-K_{\ud}({\bf r}_1,{\bf r}_2;{\bf r}_3,{\bf r}_4;\omega)
\nonumber\\
&=
\sum^{\rm occ}_{\bfk,n}
\sum^{\rm unocc}_{\bfk',n'}
\frac{
\Psi_{\bfk n \downarrow}^*(\bfr_2) \Psi_{{\bf k}n  \downarrow}({\bfr}_4)
\Psi_{{\bf k}'n' \uparrow}(\bfr_1) \Psi^*_{{\bf k'}n'\uparrow}(\bfr_3)
}{\omega-(\varepsilon_{{\bf k}' n'\uparrow}-\varepsilon_{\bfk n  \downarrow})+i \delta}
\nonumber\\
&+
\sum^{\rm unocc}_{\bfk,n}
\sum^{\rm occ}_{\bfk',n'}
\frac{
\Psi^*_{\bfk n  \downarrow}(\bfr_2)\Psi_{\bfk n  \downarrow}(\bfr_4)
\Psi_{\bfk'n' \uparrow}(\bfr_1) \Psi^*_{\bfk' n' \uparrow}(\bfr_3)
}
{-\omega-(\varepsilon_{\bfk n  \downarrow}-\varepsilon_{{\bfk'} n' \uparrow})+i \delta},
\label{eq:genpol}
\end{eqnarray} 
where $\bfk,\bfk'$ are in the first Brillouin zone,
$n (n')$ is the band index summed over occupied (unoccupied) states,
$\varepsilon_{\bfk n  \uparrow}$ ($\varepsilon_{\bfk n  \downarrow}$)
is the $n$th majority (minority) band energy at $\bfk$,
and $\Psi$ is the eigenfunction of $\hamz$.

As mentioned in Ref. \cite{LR_Sasioglu}, in order to satisfy the Goldstone theorem
$\omega(\bfq)\rightarrow0$ $(\bfq\rightarrow0)$,
we need to introduce a factor $\eta$ for $R=K(1-\eta WK)^{-1}$.
In principle, the Goldstone theorem should be automatically satisfied with
the LR method since we expect that the LR method evaluates the second derivative of the total energy of the ground states.
However, our LR is not formulated to reproduce the second derivative exactly; furthermore,
QSGW is not formulated to minimize the total energy.
This simple scaling by introducing $\eta$ is a quick remedy to satisfy the theorem; their deviations from unity show the size of vertex corrections, which should be added to the interaction $W$. The calculated $\eta$ of LDA (QSGW) are 1.15 (1.19), 1.41 (1.87), 1.26 (1.33), and 1.05 (0.87) for Fe, Ni, Co, and FeCo, respectively. These $\eta$ are in good agreement with previous calculations 1.28, 1.5, and 1.33 for Fe \cite{LR_Friedrich}, Ni \cite{LR_Sasioglu}, and FeCo \cite{LR_Friedrich}.
The deviations are not small enough.
We may need to treat the vertex correction
accurately in order to override the ambiguity due to this quick remedy in the future.

\subsection{Wannier representation}
Based on Refs. \cite{MLWF1,MLWF2},
we generate MLWFs from eigenfunctions of LDA or QSGW.
Once we generate MLWFs,
we can obtain the Wannier representation of $R^{\ud}$ as follow.

In the Wannier basis,
we expand eigenfunctions as
\begin{eqnarray} 
  \Psiknr&=\sum_{\bfR i}\akni\wkir,
  \label{eq:LCAO-lf}
\end{eqnarray} 
where $\akni$ is the expansion coefficient,
$\bfR$ is atomic position in a primitive cell,
$i$ is the Wannier orbital ($e.g.$ $i=3\dxy$) of each atom on $\bfR$.
$\wkir$ is represented as a complete set of orthogonal basis
\{$w_{\bfR i}(\bfr)$\},
\begin{eqnarray}
\wkir=\frac{1}{\sqrt{\nbvk}}\sum_{\bfT} \wirRt \exp(i\bfk\cdot\bfT),
\label{eq:wannier_B}
\end{eqnarray}
where $\bfT$ is the lattice translation vector
and $\nbvk$ is the normalization constant satisfying the Born von Karman boundary condition.
By using the orthogonality,
the eigenvalue equations $\smH\Psiknr=\ekn \Psiknr$ can be rewritten with this Wannier representation,
\begin{eqnarray}
  \sum_{\bfR' j}H_{\bfR i \bfR' j}^{\bfk}\aknj &=& \ekn \akni ,
  \label{eq:main}
\end{eqnarray}
where the Hamiltonian matrix with Wannier basis $H_{\bfR i \bfR' j}^{\bfk}$ is the Fourier transform of
$H_{\bfR i \bfR' j}^{\bfT-\bfT'} \equiv \bra{\wirRt}\smH\ket{\wjrRtp}$.

Substituting Eqs. (\ref{eq:LCAO-lf}) and (\ref{eq:wannier_B}) to \req{eq:genpol}
and using Fourier transform of real-space,
we will obtain the time-ordered linear response function for a non-interacting system
represented in a restricted Hilbert space,
\begin{eqnarray} 
&-K^{\ud}_{\bfR ij,\bfR' kl}(\bfq,\omega)\nonumber\\
&=
\frac{1}{N}\,\sum^{\rm BZ}_{\bfk}  \sum^{\rm occ}_{n } 
\sum^{\rm unocc}_{n'}\frac{\aqijkl}
{\omega-(\varepsilon_{{\bfq+\bfk} n'\uparrow}-\varepsilon_{\bfk n \downarrow})+i \delta}
\nonumber\\&
+\frac{1}{N}\,
\sum^{\rm BZ}_{\bfk} \sum^{\rm  unocc}_{n } \sum^{\rm occ}_{n'}
\frac{\aqijkl}
{-\omega-(\varepsilon_{\bfk n \downarrow}-\varepsilon_{{\bfq+\bfk} n' \uparrow})+i \delta}.
\label{eq:imk_wan}
\end{eqnarray} 

We calculate the imaginary part of
$-K^{\ud}_{\bfR ij,\bfR' kl}(\bfq,\omega)$ by a tetrahedron method 
and obtain its real part by the Hilbert transform. 
The matrix element of $R^{\ud}_{{\bf R}ij,{\bf R'}kl}$ is calculated 
through $R=K(1-\eta WK)^{-1}$, where $W$ is calculated in the random phase approximation (RPA)
in the product basis technique developed in Ref. \cite{Kotani_2001}.

\subsection{Calculation details}
All of the calculation procedures above are implemented in the
first-principles package $ecalj$ \cite{ecalj,Deguchi_2016}.
The $ecalj$ is based on the linearized augmented
plane-wave and muffin-tin orbital (MTO) method (PMT method),
which combines augmented plane wave (APW) and MTO basis sets.
We also generate MLWFs in $ecalj$.
We perform LDA and QSGW calculations for band structures
with $20\times20\times20$ and $16\times16\times16$ $k$-point mesh respectively.
We consider 9 MLWFs ($spd$) for the $3d$ elemental materials (Fe and Ni) and 18 MLWFs for hcp Co and binary FeCo.
In the calculations of $K^{\ud}$,
we use $48\times48\times48$ $q$-point mesh for the 3$d$ elemental material
and $24\times24\times24$ for binary FeCo.
We use static and onsite $W$, $i.e.$, we take $W_{ijkl}(\omega)=W_{\bfR ij,\bfR kl}(\omega=0)$.
We use experimental lattice parameters, $a=2.867$ \AA, $a=3.524$ \AA, $a=2.850$ \AA\
for Fe, Ni, and FeCo, respectively.
For hcp Co, we use $a=2.507$ \AA\ and $c=4.070$ \AA.

\section{Results and discussion}
\subsection{bcc Fe}
\label{bccfe}
\begin{figure}[t]
\includegraphics[width=\linewidth]{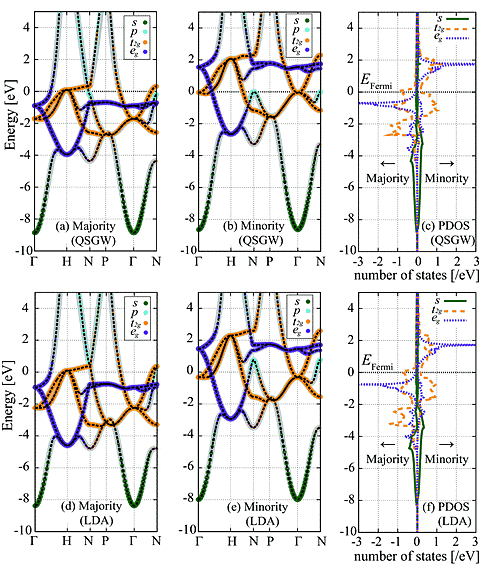}
\caption{Calculated band structures of Fe in QSGW ((a) majority spin, (b) minority spin) 
and in LDA ((d) majority , (e) minority spin).
The interpolated bands based on 9 MLWFs are also shown (broken line) with original bands (bold gray line).
Size of colored circles on the bands shows the weight of MLWF bands.
Partial density of states for $4s$, $t_{2g}$, and $e_g$ in QSGW and LDA are shown in (c) and (f). 
Fermi energy $\ef$ is set to 0 eV.
}
\label{f1}
\end{figure}

\refigs{f1}(a), (b), and (c) show the majority and minority band structures 
and the partial density of states in QSGW for Fe,
while \refigs{f1}(d), (e), and (f) in LDA as well.
Calculated total magnetic moments in LDA and QSGW are both 2.22 $\mb$
for Fe, in agreement with the experimental value 2.22 $\mb$ \cite{mmom_Danan},
in contrast to 2.93 $\mb$ in the fully self-consistent $GW$ method \cite{scGW}.
Our results are consistent with Ref. \cite{Sponza} by Sponza {\it et al}.
The superposed Wannier band structures in \req{eq:main} by broken lines
are entirely on the original band structures by bold grey lines.
Size of colored circles show the weights of each MLWF. 
In \retable{arpes}, we show the $t_{2g}$ of minority spin at $\Gamma$ and that of majority spin at N in LDA and QSGW. 
QSGW gives better agreement with the
angle-resolved photoemission spectroscopy (ARPES) data \cite{ARPES_Fe}. 
The $3d$-bandwidth in QSGW is a little smaller than that in LDA.
Except for this difference, the overall shapes of the majority and the minority bands are similar in both LDA and QSGW.

\begin{figure}[b]
\includegraphics[width=\linewidth]{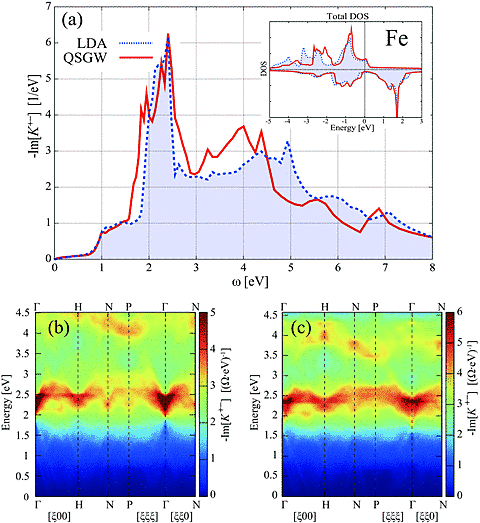}
\caption{(a) Calculated $\imko$ in Fe
in QSGW (red bold line) and in LDA (blue broken line).
The inset is the total density of states in Fe.
(b) and (c) show calculated $-{\rm Im}{[K^{+-}({\bf q},\omega)}]$ along the BZ symmetry line
in LDA and QSGW, respectively.
$\Omega$ is the unit cell volume.
}
\label{f2}
\end{figure}

\begin{table}[t!]
\begin{center}
  \caption{$t_{2g}$ level of Fe at $\Gamma$ for the minority spin,
    and that at N for the majority spin. These are in LDA, in QSGW,
    in addition to the experimental data by ARPES \cite{ARPES_Fe}. Energy is relative to $\ef$.}
\label{arpes}
\begin{tabular}{c@{\hspace{5mm}}c@{\hspace{5mm}}c@{\hspace{5mm}}c@{\hspace{5mm}}c}
  \hline\hline
  &\multicolumn{3}{c}{band energy [eV]}\\\cline{2-4}
  &LDA&QSGW&Expt. \cite{ARPES_Fe}\\\hline
  $\Gamma$(Minority)&-0.32&-0.11&-0.19\\
  N (Majority)&-0.74&-0.68&-0.57\\
  \hline\hline
\end{tabular}
\end{center}
\end{table}

\refig{f2}(a) shows $\imko$ in LDA and in QSGW, where
$K^{+-}(\bfq, \omega)$ means the trace of the matrix
$K^{\ud}$ given as $K^{+-}(\bfq,\omega) = \sum_{\bfR,i,j} K^{\ud}_{\bfR ii, \bfR jj}(\bfq,\omega)$.
We use a little different definition from
Refs. \onlinecite{LR_Sasioglu,LR_Friedrich,LR_Friedrich2018},
thus it is not meaningful to compare absolute value of $K^{+-}(\bfq,\omega)$ with their results.
As shown in the figure, QSGW gives smaller $\dex$ and 3$d$-bandwidth,
which is consistent with results by Sponza {\it et al}.
Roughly speaking, the shape of $-{\rm Im}[K^{+-}(\bfq=0,\omega)]$ agree with the shape
of density of states (DOS) of majority spin. The peak around 2 eV originates from the $t_{2g}^{\uparrow}$-$t_{2g}^{\downarrow}$
and $e_g^{\uparrow}$-$e_g^{\downarrow}$ transition,
{\it i.e.}, vertical transitions to the unoccupied minority states above the Fermi energy $\ef$
from the occupied majority states just below the $\ef$ in \refig{f1}.
The second peak around 4 eV is stemmed from
another $e_g^{\uparrow}$-$e_g^{\downarrow}$ transition to $\ef+2$ eV in minority states from $\ef-2$ eV in majority states.

We see two features in the difference between LDA and QSGW in $\imko$
shown in \refig{f2}(a). One is that the width of the peak around $2$ eV in QSGW is wider than that
in LDA. The difference of DOS in LDA and QSGW can not explain this
fact; it can be due to the difference of eigenfunctions.
The peak becomes wider in QSGW, probably because of the
general tendency of QSGW that it makes a more significant difference between
occupied $3d$ states and unoccupied $3d$ states. The former is more localized, and the latter more extended in comparison with the case
in LDA. The other is the width due to the $3d$ band; corresponding to
the width of $3d$ band shown in the inset of \refig{f2}(a), we see narrower width in $\imko$ in QSGW.

\refigs{f2}(b) and (c) show the Stoner excitation spectrum $-{\rm Im}[K^{+-}(\bfq,\omega)]$ in LDA and QSGW.
Our LDA results give good agreement with Fig. 6 in Ref. \onlinecite{LR_Friedrich2018}.
We see red triangle-like strong intensity around $\Gamma$, especially in LDA.
The center of peak moves up as a function of $\bfq$.
This is because shifted $q$ from $\Gamma$ requires corresponding energy shift to trace the peak of $-{\rm Im}[K^{+-}(\bfq,\omega)]$ as a function of $\omega$.
This is explained in Fig. 7 of Ref. \onlinecite{LR_Friedrich2018}.

\begin{figure}[t]
\includegraphics[width=\linewidth]{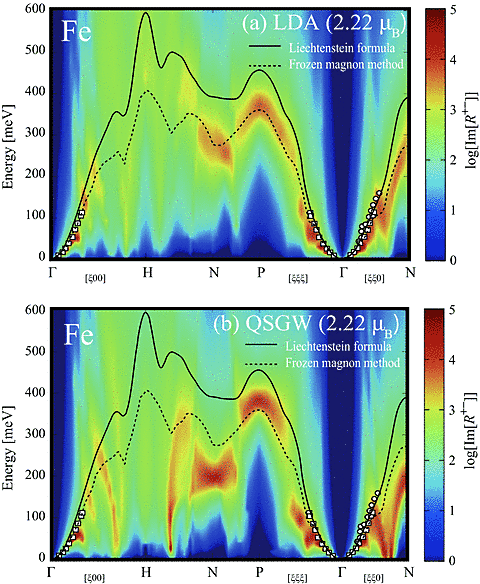}
\caption{
$\ImRpm$ for Fe (a) in LDA and (b) in QSGW, showing the SW dispersion;
we see slight discontinuities because of the mesh of used $\bfk$ points.
Results with LF \cite{Pajda} (solid line), and
that with FMM \cite{FMM_Halilov} (broken line) are superposed. 
Experimental data by neutron scattering are indicated by open squares (Fe (12\%Si) at RT \cite{Fe_expt1}) and open circles (pure Fe at 10 K \cite{expt_Loong}.) 
}
\label{f3}
\end{figure}

\begin{table*}[t!]
\begin{center}
\caption{Calculated stiffness constant $D$ for Fe, Ni, Co and FeCo.
The results by other groups are shown together; the LR \cite{LR_Friedrich},
with the LF \cite{Pajda},
and with the time-dependent DFT (TDDFT) \cite{Niesert} (on average).
In addition we show inelastic neutron scattering data
\cite{Fe_expt1,Fe_expt2,Ni_expt2,Ni_expt3,hcpCo_Shirane,hcpCo_Shirane,feco_Lowde}.}
\label{stiffness}
\scalebox{1}{
\begin{tabular}{c@{\hspace{3mm}}c@{\hspace{3mm}}c@{\hspace{3mm}}c@{\hspace{3mm}}c@{\hspace{3mm}}c@{\hspace{3mm}}cc}
  \hline\hline
  &\multicolumn{6}{c}{$D$ [meV$\cdot$\AA$^2$]}\\
  \cline{2-7}
  Material&LR (LDA)&LR (QSGW)&Expt.&LR (GGA) \cite{LR_Friedrich}&LF \cite{Pajda}&TDDFT \cite{Niesert}\\
\hline
bcc Fe& 155& 222& 230 (RT) \cite{Fe_expt1}&
248& 250& 189\\
& & & 280 (4.2 K) \cite{Fe_expt2}\\
fcc Ni& 873& 449 & 433 \cite{Ni_expt2}& & 756& 1097\\
& & &555 \cite{Ni_expt3}\\
hcp Co [100]& 565 & 486 & 478 \cite{hcpCo_Perring}\\ 
hcp Co [001]& 752 & 532 & 410 \cite{hcpCo_Perring}\\
& & & 510 \cite{hcpCo_Shirane}\\
B2 FeCo & 407 & 307 & 450-500 \cite{feco_Lowde}\\
\hline\hline
\end{tabular}
}
\end{center}
\end{table*}

\refig{f3} shows $\ImRpm$ in LDA (a) and in QSGW (b), where $R^{+-}(\bfq,\omega)$ means the trace of the matrix $R^{\ud}$ given as $R^{+-}(\bfq,\omega) = \sum_{\bfR,i,j} R^{\ud}_{\bfR ii, \bfR jj}(\bfq,\omega)$.
We superpose experimental data \cite{Fe_expt1,expt_Loong} on it.
We also superpose the SW dispersion calculated with the LF
\cite{Pajda} in LDA, and that with FMM in LDA \cite{FMM_Halilov}.
These are not only in (a) but also in (b) as a guide of eye.
As shown in \refig{f3}, the peak broadening due to the Stoner damping can be seen even below 100 meV because bcc Fe is a weak ferromagnet, whose majority and minority $3d$ have
relatively large DOS at $\ef$ as shown in the inset of \refig{f2}(a).
This results in relatively large low-energy Stoner excitations.
It means that SWs are getting to be hybridized well with Stoner
excitation immediately after departing from $\Gamma$. 
The strong damping around H is also seen in the previous calculation combining the
the generalized gradient approximation (GGA) and the MLWF approach with 6 MLWFS ($sd$) \cite{LR_Friedrich}.
Our LDA calculation indicates Kohn anomalies in $\Gamma$-H, H-N, and $\Gamma$-N, which are also found in the other calculations \cite{Pajda, FMM_Halilov, Grotheer}.
We checked calculations with denser q-point mesh (60$\times$60$\times$60)  and confirmed the strong anomaly at 2/3 along $\Gamma$-N in LDA, and especially in QSGW.
Ref. \cite{LR_Friedrich2018} explains how such anomalies can be
traced back to the band structures, although they have not given
explicit analysis. Real metals such as Fe can have complicated
band structures, resulting in too complicated Fermi-surface-nestings like phenomena to be analyzed. 
Thus, we also have not yet got into such analysis. We are somehow
skeptical whether it is worth to do or not.


In \retable{stiffness}, we summarize calculated results of stiffness constant $D$,
with another LR result based on the GGA \cite{LR_Friedrich},
and with that of the time-dependent density functional theory (TDDFT) \cite{Niesert}. 
To obtain $D$, we fit the calculated SW dispersion by quadratic functions. 
For the fitting, we just take peaks for small $q$ as $|{\bf q}|<0.20(\frac{2\pi}{a})$ 
where little Stoner damping occurs.
Details for Fe and Ni are in supplements \cite{Supplemental}.
LDA gives $D=155$ meV$\cdot$\AA$^2$, which is a little smaller than experiments
$D=230$, $280$ meV$\cdot$\AA$^2$ \cite{Fe_expt1,Fe_expt2}.
On the other hand, QSGW gives $D=222$ meV$\cdot$\AA$^2$ in much better
agreement with the experimental values.
Note that we see a contradiction between our LR (LDA) and the other two
previous calculations, the LR (GGA) and the LF.
Our values $D=155$ meV$\cdot$\AA$^2$ is too low in comparison with the other data 248, 250 meV$\cdot$\AA$^2$, although the smaller difference from $D=189$ meV$\cdot$\AA$^2$ in TDDFT.
However, we currently have no definite idea to resolve the discrepancy from these previous works.

\subsection{fcc Ni}
\begin{figure}[b!]
\includegraphics[width=\linewidth]{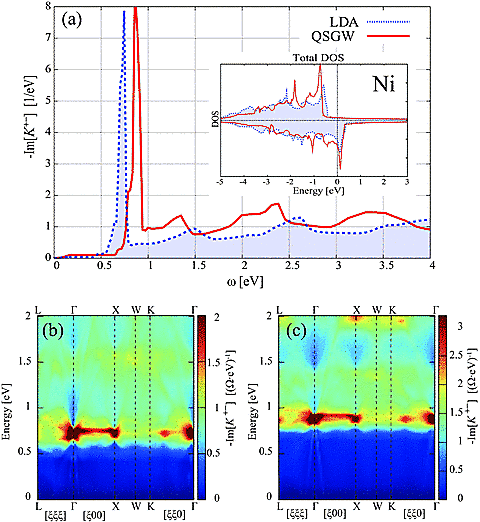}
\caption{
(a) $\imko$ in Ni in QSGW (red bold line) and in LDA (blue broken line).
  The inset is the total density of states in Ni.
  (b) and (c) calculated $-{\rm Im}[{K^{+-}({\bf q},\omega)}]$ along the BZ symmetry line
in LDA and QSGW, respectively.
$\Omega$ is the unit cell volume.
}
\label{f4}
\end{figure}

\begin{figure}[b]
\includegraphics[width=\linewidth]{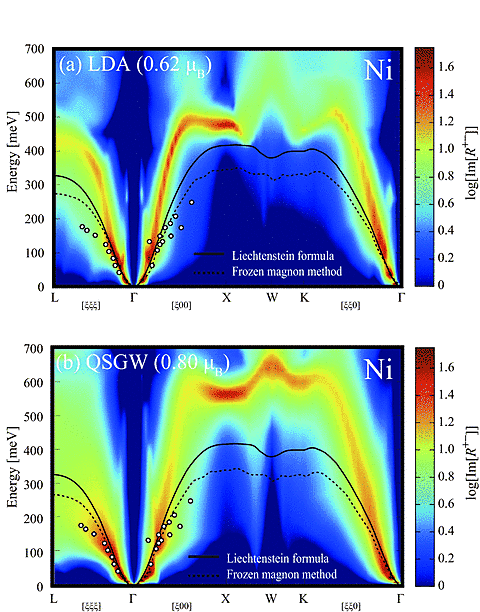}
\caption{
$\ImRpm$ for Ni in LDA (a) and in QSGW (b), showing the SW dispersion.
We superpose other results with the LF \cite{Pajda} (solid line) and with 
FMM \cite{FMM_Halilov} (broken line).
Experimental results by neutron scattering \cite{Ni_expt4} are indicated by circles. 
}
\label{f5}
\end{figure}

The calculated magnetic moment for Ni in LDA is in agreement with the experiment, 0.62 $\mb$ \cite{mmom_Danan}. On the other hand, QSGW gives 0.80 $\mb$.
Sponza {\it et al}. \cite{Sponza} indicates
that this is reasonable because we have not taken into account the
longitudinal quantum spin fluctuation. In LDA, we may have accidentally had a good agreement because
of too small exchange splitting cancels the fact that calculations do not include the fluctuation.

\refig{f4}(a) shows the $\imko$ in Ni.
Peaks at 0.7 eV and 0.8 eV in LDA and QSGW are the Stoner gaps,
corresponding to the difference of peaks between majority and minority
spins in DOS shown in its inset.
$\dex$ given in LDA and QSGW are about two times
larger than 0.3 eV, which is the value obtained by ARPES at $L_3$ point \cite{exchange_Ni}.
Sponza {\it et al}. \cite{Sponza} indicates
that the overestimation is due to the missing of spin fluctuations.
\refigs{f4}(b) and (c) show $-{\rm Im}[K^{+-}(\bfq,\omega)]$ in LDA and
QSGW. 
Our LDA results give good agreement with Fig. 6 of Ref. \onlinecite{LR_Friedrich2018}.
We see that strong intensity around $\Gamma$ get broadened 
as a function of $\bfq$ as in the case of homogeneous electron gas shown in
Fig. 5 of Ref. \onlinecite{LR_Friedrich2018}. In QSGW, 
$\bfq$-dependence of $-{\rm Im}[K^{+-}(\bfq,\omega)]$ looks slightly weakened
around $\Gamma$, probably because of the reflection of flattened (weak $q$-dependent) 3$d$ band.

\begin{figure}[b!]
\includegraphics[width=\linewidth]{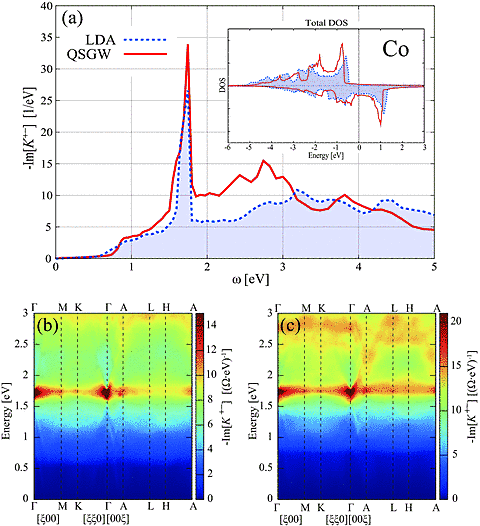}
\caption{
(a) $\imko$ in Co in QSGW (red bold line) and in LDA (blue broken line).
The inset is total density of states in Co.
(b) and (c) show calculated $-{\rm Im}[{K^{+-}({\bf q},\omega)}]$ along the BZ symmetry line
in LDA and QSGW, respectively.
$\Omega$ is the unit cell volume.
}
\label{f4_2}
\end{figure}

\begin{figure}[b!]
\includegraphics[width=\linewidth]{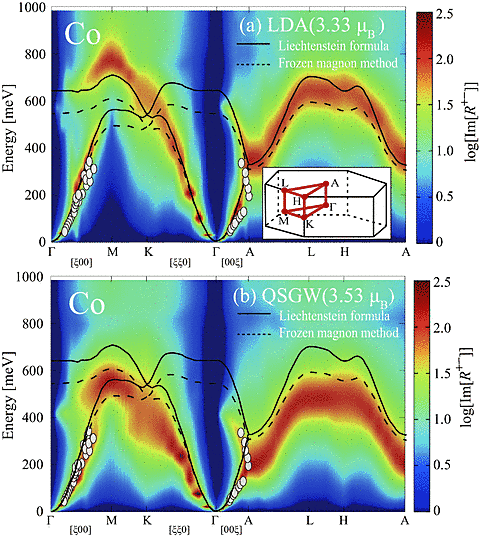}
\caption{
$\ImRpm$ for Co in LDA (a) and in QSGW (b), showing the SW dispersion.
The LF \cite{Pajda} (bold line),
the FMM calculation \cite{FMM_Halilov} (broken line) are also shown.  
Experimental data by neutron scattering \cite{hcpCo_Perring} are indicated by circles. 
The inset shows the BZ for hcp Co and its symmetry lines.
}
\label{sw_co}
\end{figure}

In \refig{f5} (a), we show $\ImRpm$ in LDA.
We can identify the SW dispersion in the whole BZ
in contrast to the case of Fe in \refig{f3}.
Our SW dispersion in LDA is consistent with a previous LR calculation
by Savrasov \cite{LR_Savrasov} and a TDDFT calculation by Niesert \cite{Niesert}.
As superposed in \refig{f5},
results with FMM \cite{FMM_Halilov} and with the LF \cite{Pajda} 
give a little lower $\SWE$.
Let us compare QSGW result shown in \refig{f5}(b) 
with (a), where we can use black lines as a guide of eye.
$\SWE$ curvature around $\Gamma$ is smaller in QSGW. 
In fact, \retable{stiffness} shows that
QSGW gives very smaller $D=449$ meV$\cdot$\AA$^2$ around $\Gamma$
than $D=873$ meV$\cdot$\AA$^2$ in LDA. This is in agreement with
the experimental values $D=433$, $555$ meV$\cdot$\AA$^2$ \cite{Ni_expt2,Ni_expt3}.
This is the reflection of weak $\bfq$-dependence of
$-{\rm Im}[K^{+-}(\bfq,\omega)]$ around $\Gamma$ in the previous paragraph.
Along $\Gamma$-L, QSGW successfully trace an experiment \cite{Ni_expt4}
even up to the half of the BZ boundary.
Although (b) may be taken as a simple elongation of (a) at a glance,
it is not true if we take the behavior around $\Gamma$ into account.
In Ref. \cite{LR_Karlsson}, Karlsson and Aryasetiawan gives good
agreement with the SW dispersion along [100] by adjusting the $\dex$ of Ni.
However, such a procedure may give a simple shrinkage.
Thus the physical mechanism in QSGW is very different from their method
even though both our QSGW and their method reproduce the experimental $D$.


\subsection{hcp Co}
\refig{f4_2}(a) shows the $\imko$ in Co and
\refigs{f4_2}(b) and (c) show $-{\rm Im}[K^{+-}(\bfq,\omega)]$ in LDA and
QSGW.
The calculated magnetic moments per Co atom is 
1.67 $\mb$ in LDA, 1.76 $\mb$ in QSGW. 
These are a little larger than the experiment 1.58 $\mb$ \cite{hcpCo_moment}.
It is reasonable in the sense that the QSGW value relative to
experiment is 1.76 $\mb$/1.58 $\mb$, in between 2.22 $\mb$ /2.22 $\mb$ (Fe) and 0.80 $\mb$ /0.62 $\mb$  (Ni).
Let us compare peaks of $3d$ shown in insets with those for Fe and Ni (Figs. \ref{f2} and \ref{f4}).
In QSGW, $3d$ bands are narrower than LDA in both of majority, and minority spins in Co and Ni, in contrast to the case of Fe where little narrowing of DOS in the minority spins. It is probably because 
the bcc structure has more hybridization with $sp$ bands than fcc and hcp.
In Co, the largest peaks of $3d$ are pushed down by QSGW relative to LDA, with keeping the exchange splitting. Thus changes of $\imko$ from
QSGW to LDA are similar in Fe and Co. As we already noted in Sec.\ref{bccfe},
we admit several universal tendencies of QSGW relative to LDA, however,
such changes of DOS and $\imko$ are hardly predicted without calculations in practice.

In \refig{sw_co}(a), we show $\ImRpm$ in LDA together with 
plots of the SW dispersion given by the FMM \cite{FMM_Halilov} (black broken lines)
and by the LF \cite{Pajda} (black lines).
In these plots, two branches appear because of two atoms per primitive cell.
The LF traces peaks of our $\ImRpm$ very
well especially along $\Gamma$-A-K-H-A.
At M around, the black lines are slightly lower than the peak of $\ImRpm$ seen at $\sim$800 meV. Near $\Gamma$, $\ImRpm$ shows no optical branch.
Experimental data shown by oval circles \cite{hcpCo_Perring,hcpCo_Shirane} are a little lover than the plots and peaks of $\ImRpm$.

In contrast, we have an impressive agreement with the experiment in QSGW.
As seen in \refig{sw_co}(b), oval circles are on the peak of $\ImRpm$ in QSGW.
The calculated $D$ shown in \retable{stiffness} in QSGW are
486 meV$\cdot$\AA$^2$ along [100], and 532 meV$\cdot$\AA$^2$ along [001].
These give much better agreements with experiments, consistent
with the agreement in \refig{sw_co}(b).
This agreement of the SW energy is probably originated from narrower $3d$ band in QSGW,
resulting weaker ${\bf q}$-dependence of $-{\rm Im}[K^{+-}({\bf q},\omega)]$, rather than LDA.

\subsection{B2 FeCo}
\begin{figure}[b!]
\includegraphics[width=\linewidth]{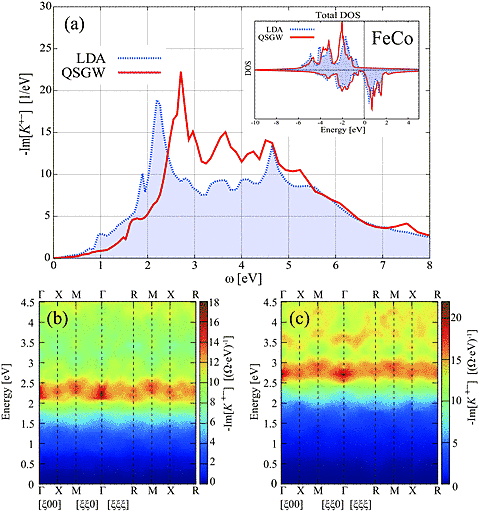}
\caption{
(a) $-{\rm Im}[K^{+-}(\bfq=0,\omega)]$ of FeCo
in QSGW (red bold line) and LDA (blue broken line).
The inset is the total density of states in FeCo.
(b) and (c) show calculated $-{\rm Im}[{K^{+-}({\bf q},\omega)}]$ along the BZ symmetry line
in LDA and in QSGW, respectively.
$\Omega$ is the unit cell volume.
}
\label{imk_feco}
\end{figure}

We treat B2 FeCo in the CsCl structure.
Calculated magnetic moments per cell are
4.44 $\mb$ in LDA, 4.80 $\mb$ in QSGW.
The latter is close to experiment 4.70 $\mb$ \cite{feco_Goldman}.
It is consistent with other compounds \cite{TMO_QSGW,LSMO_QSGW} where QSGW give agreements with experiments as for magnetic moments when LDA gives underestimation. Alternatively, we may take FeCo as a case between Fe and Co. Since QSGW/experiment = 2.22 $\mb$/2.22 $\mb$ for Fe,
= 1.76 $\mb$/1.58 $\mb$ for Co, we may say that slight overestimation 4.80 $\mb$/4.70 $\mb$ is reasonable.

\refig{imk_feco}(a) shows $\imko$ in LDA and QSGW.
In its inset, $\dex$ is $\sim$ 2.8 eV in QSGW while $\sim$2.2 eV in LDA. The difference results in the difference of peaks in $\imko$.
\refigs{imk_feco}(b) and (c) show $-{\rm Im}[K^{+-}(\bfq,\omega)]$ in LDA and
QSGW, although we see no specific features worth to be mentioned.

\begin{figure}[b!]
\includegraphics[width=\linewidth]{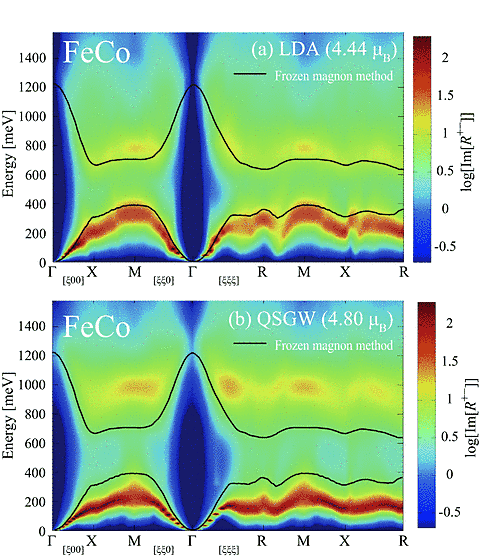}
\caption{
$\ImRpm$ for FeCo (a) in LDA and (b) in QSGW, showing the SW dispersion.
The black bold line shows the FMM result \cite{Grotheer} in LDA.
}
\label{sw_feco}
\end{figure}

\refig{sw_feco} shows $\ImRpm$ in (a) LDA and in (b) QSGW,
together with the previous SW calculation in the FMM \cite{Grotheer}.
$\ImRpm$ in LDA shows the lower peaks of $\SWE$ than FMM.
$\ImRpm$ in LDA gives $D=407$ meV$\cdot$\AA$^2$ is a little smaller than
500 meV$\cdot$\AA$^2$ by Grotheer \cite{Grotheer}.
The optical branch is weakened as in the case of Fe.
Weak peak around $\sim$ 600 meV are close to $\SWE$ in FMM.

In QSGW, there is lower $\SWE$ in the whole BZ as in the case of Co.
\retable{stiffness} shows that 
$D=307$ meV$\cdot$\AA$^2$ in QSGW is much smaller than the experiment
450-500 meV$\cdot$\AA$^2$ by inelastic neutron scattering \cite{feco_Lowde}.
Considering success on Fe, Ni, and Co, this FeCo was the case that we could expect a good agreement with experiments. We have not yet found a reason
why QSGW gives such discrepancy from the experiment.


\section{Summary}
In order to calculate SW dispersion in QSGW, 
we have implemented an effective numerical method for
calculating $\chipm$ in a package $ecalj$.
This is in the linear response formulation based on the maximally
localized Wannier functions as given in Ref. \onlinecite{LR_Sasioglu}.

Then we apply the method to Fe, Ni, Co, and FeCo.
We compare peak of $\ImRpm$ with inelastic neutron scattering data and with the spin stiffness $D$.
For Fe, Ni, and Co, QSGW gives much better agreements with the experiment
rather than LDA does. Notably, too large $D$ of Ni in LDA is reduced
by half, resulting in a good agreement with the experiment. We see similar
agreement for Co in comparison with the neutron scattering data.
For FeCo, we have not yet understood why $D$ in QSGW disagree with the experiment.

Such good agreements are owing to the reliable description of
the electronic structure in QSGW. QSGW gives a good description of $3d$-bandwidth, $\dex$ and magnetic moments,
except the case of Ni where we have a too large magnetic moment.
Our method developed here is promising in the sense that it covers wide
range of materials from metals treated here to transition-metal oxides
where LDA can be hardly applicable.

\begin{acknowledgments}
This work was partly supported by the Building of Consortia for the Development of Human Resources in Science and Technology project,
implemented by the Ministry of Education, Culture, Sports, Science, and Technology (MEXT) of Japan.
This work was partly supported by JST CREST Grant number JPMJCR1812 and by JSPS KAKENHI Grant Number JP18H05212.
T. Kotani thanks to supporting by JSPS KAKENHI Grant Number 17K05499.
We also thank the computing time provided by Research Institute for Information Technology (Kyushu University).
We want to thank T. Fukazawa for giving us useful comments.
\end{acknowledgments}



\begin{thebibliography}{9}
\bibitem{Fe_expt1} 
  J. W. Lynn,
  {\it Phys. Rev. B} {\bf 11}, 2624 (1975).
\bibitem{Ni_expt2}
  H. A. Mook, R. M. Nicklow, E. D. Thompson, and M. K. Wilkinson,
  {\it J. Appl. Phys.} {\bf 40}, 1450 (1969).
\bibitem{lsmo_expt}
  F. Ye, P. Dai, J. A. Fernandez-Baca, D. T. Adroja,
  T. G. Perring, Y. Tomioka, and Y. Tokura,
  {\it Phys. Rev. B} {\bf 75}, 144408 (2007).
\bibitem{Stoner_Vollmer}
  R. Vollmer, M. Etzkorn, P. S. A. Kumar, H. Ibach, and J. Kirschner,
  {\it Thin Solid Films} {\bf 464}, 42 (2004).
\bibitem{Bloch_rule}
  C. Kittel,
  {\it Introduction to Solid State Physics}, 8th ed.
  (John Wiley and Sons Ltd, 2004) p. 335.
\bibitem{Fe_expt2}
  R. Pauthenet,
  {\it J. Appl. Phys.} {\bf 53}, 8187 (1982).
\bibitem{Lichtenstein}
  A. I. Lichtenstein, M. I. Katsnelson, V. P. Antropov,
  and V. A. Gubanov,
  {\it J. Magn. Magn. Mater.} {\bf 67}, 65 (1987).
\bibitem{MFT}
  A. Oswald, R. Zeller, P. J. Braspenning, and P. H. Dederichs,
  {\it J. Phys. F: Met. Phys.} {\bf 15}, 193 (1985).
\bibitem{Pajda}
  M. Pajda, J. Kudrnovsk\'y, I. Turek, V. Drchal and P. Bruno,
  {\it Phys. Rev. B} {\bf 64}, 174402 (2001).
\bibitem{FMM_Halilov}
  S. V. Halilov, H. Eschrig, A. Y. Perlov, and P. M. Oppeneer
  {\it Phys. Rev. B} {\bf 58}, 293 (1998).
\bibitem{Grotheer} 
  O. Grotheer, C. Ederer, and M. F\"{a}hnle,
  {\it Phys. Rev. B} {\bf 63}, 100401(R) (2001).
\bibitem{LR_Gross}
  E. K. U. Gross and W. Kohn,
  {\it Phys. Rev. Lett.} {\bf 55}, 2850 (1985).
\bibitem{LR_Cooke}
  J. F. Cooke, J. W. Lynn, and H. L. Davis,
  {\it Phys. Rev. B} {\bf 21}, 4118 (1980).
\bibitem{LR_Savrasov}
  S. Y. Savrasov,
  {\it Phys. Rev. Lett.} {\bf 81}, 2570 (1998).
\bibitem{LR_Karlsson}
  K. Karlsson and F. Aryasetiawan,
  {\it Phys. Rev. B} {\bf 62}, 3006 (2000).
\bibitem{LR_Sasioglu}
  E. \c{S}a\c{s}{\i}o\v{g}lu, A. Schindlmayr, C. Friedrich, F. Freimuth, and S. Bl\"{u}gel,
  {\it Phys. Rev. B} {\bf 81}, 054434 (2010).

\bibitem{Sponza}
  L. Sponza, P. Pisanti, A. Vishina, D. Pashov, C. Weber, M. van Schilfgaarde,
  S. Acharya, J. Vidal, and G. Kotliar,
  {\it Phys. Rev. B} {\bf 95}, 041112(R) (2017).
\bibitem{TMO_QSGW}
  T. Kotani and M. van Schilfgaarde,
  {\it J. Phys. Condens. Matter} {\bf 20}, 295214 (2008).
\bibitem{LSMO_QSGW}
  T. Kotani and H. Kino,
  {\it J. Phys. Condens. Matter} {\bf 21}, 266002 (2009).
\bibitem{Deguchi_2016}
  D. Deguchi, K. Sato, H. Kino, and T. Kotani,
  {\it Jpn. J. Appl. Phys.} {\bf 55}, 051201 (2016).
\bibitem{ecalj}
  A first-principles electronic-structure suite based on the PMT method, ecalj package,
  is freely available at https://github.com/tkotani/ecalj.
  Its one-body part is developed based on the LMTO part in the LMsuit package at http://www.lmsuite.org/.
  
\bibitem{hedin_gw1}
  L. Hedin,
  {\it Phys. Rev. A} {\bf 139}, 796 (1965).
\bibitem{hedin_gw2}
  M. S. Hybertsen and S. G. Louie,
  {\it Phys. Rev. Lett.} {\bf 55}, 1418 (1985).
\bibitem{HybersteinLouie1986}
  M. S. Hybertsen and S. G. Louie,
  {\it Phys. Rev. B} {\bf 34} 5390 (1986).
\bibitem{qsgw1}
  M. van Schilfgaarde, T. Kotani, and S. Faleev,
  {\it Phys. Rev. Lett.} {\bf 96}, 226402 (2006).
\bibitem{qsgw2}
  T. Kotani,
  {\it J. Phys. Soc. Jpn.} {\bf 83}, 094711 (2014).
\bibitem{qsgw3}
  T. Kotani, H. Kino, and H. Akai,
  {\it J. Phys. Soc. Jpn.} {\bf 84}, 034702 (2015).
\bibitem{LR_Friedrich}
  C. Friedrich, E. \c{S}a\c{s}{\i}o\v{g}lu, M M\"{u}ller, A. Schindlmayr, and S. Bl\"{u}gel,
  {\it Spin Excitations in Solids from Many-Body Perturbation Theory.}
  (In: Di Valentin C., Botti S., Cococcioni M. (eds)
  {\it First Principles Approaches to Spectroscopic Properties of Complex Materials.}
  Springer, Berlin, Heidelberg, 2014) p. 259. 


\bibitem{MLWF1}
  N. Marzari and D.Vanderbilt,
  {\it Phys. Rev. B} {\bf 56}, 12847 (1997).
\bibitem{MLWF2}
  I. Souza, N. Marzari and D.Vanderbilt,
  {\it Phys. Rev. B} {\bf 65}, 035109 (2001).
\bibitem{Kotani_2001}
  T. Kotani and M. van Schilfgaarde, {\it Solid State Commun.} {\bf 121}, 461 (2002).
\bibitem{mmom_Danan}
  H. Danan, A. Herr, and A. J. P. Meyer,
  {\it J. Appl. Phys.} {\bf 39}, 669 (1968).
\bibitem{scGW}
  A. L. Kutepov,
  {\it J. Phys. Condens. Matter} {\bf 29}, 465503 (2017).
\bibitem{ARPES_Fe}
  J. Sch\"{a}fer, M. Hoinkis, E. Rotenberg, P. Blaha, and R. Claessen,
  {\it Phys. Rev. B} {\bf 72}, 155115 (2005).
\bibitem{LR_Friedrich2018}
  C. Friedrich, M. C. T. D M\"{u}ller, S. Bl\"{u}gel
  {\it Spin Excitations in Solid from Many-Body Perturbation Theory.}
  (In: Andreoni W., Yip S. (eds) {\it Handbook of Materials Modeling.} Springer, Cham, 2018) p. 1. 

\bibitem{expt_Loong}
  C. K. Loong, J. M. Carpenter, J. W. Lynn, R. A. Robinson, and H. A. Mook,
  {\it J. Appl. Phys.} {\bf 55}, 1895 (1984).

\bibitem{Niesert}
  M. Niesert,
  {\it Ab initio Calculations of Spin-Wave Excitation Spectra from Time-Dependent Density-Functional Theory},
  (Ph.D. thesis, RWTH Aachen University, 2011).
\bibitem{Supplemental}
  See Supplemental Materials for detailed fitting results of SW dispersion.

\bibitem{exchange_Ni}
  D. E. Eastman, F. J. Himpsel, and J. A. Knapp,
  {\it Phys. Rev. Lett.} {\bf 44}, 95 (1980).

\bibitem{Ni_expt3}
  H. A. Mook, J. W. Lynn, and R. M. Nicklow,
  {\it Solid State Commun.} {\bf 30}, 556 (1973).
\bibitem{Ni_expt4} 
  H. A. Mook and D. M. Paul,
  {\it Phys. Rev. Lett.} {\bf 54}, 227 (1985).
  
\bibitem{hcpCo_moment}
  H. P. Myers and W. Sucksmith,
  {\it Proc. R. Soc. A} {\bf 207}, 427 (1951).
\bibitem{hcpCo_Perring}
  T. G. Perring, A. D. Taylor, and G. L. Squires,
  {\it Physica B} {\bf 213}\&{\bf 214}, 348 (1995).
\bibitem{hcpCo_Shirane}
  G. Shirane, V. J. Minkiewicz, and R. Nathans,
  {\it J. Appl. Phys.} {\bf 39}, 383 (1968).

\bibitem{feco_Goldman}
  J. E. Goldman and R. Smoluchowski,
  {\it Phys. Rev.} {\bf 75}, 310 (1949).

\bibitem{feco_Lowde}
  R. D. Lowde, M. Shimizu, M. W. Stringfellow, and B. H. Torrie,
  {\it Phys. Rev. Lett.} {\bf 14}, 698 (1965).
\end{thebibliography}
\end{document}